%% file: main.tex
\documentclass[conference]{IEEEtran}

\input{includes}

\IEEEoverridecommandlockouts
\usepackage{cite}
\usepackage{amsmath,amssymb,amsfonts}
\usepackage{graphicx}
\usepackage{textcomp}
\usepackage{xcolor}
\def\BibTeX{{\rm B\kern-.05em{\sc i\kern-.025em b}\kern-.08em
    T\kern-.1667em\lower.7ex\hbox{E}\kern-.125emX}}

\begin{document}

\title{PEFT-as-an-Attack! Jailbreaking Language Models during Federated Parameter-Efficient Fine-Tuning}

\author{
\IEEEauthorblockN{
Shenghui Li\IEEEauthorrefmark{1}~~~~
Edith C.-H. Ngai\IEEEauthorrefmark{2}~~~~
Fanghua Ye\IEEEauthorrefmark{3}~~~~
Thiemo Voigt\IEEEauthorrefmark{1}\IEEEauthorrefmark{4}}
\IEEEauthorblockA{\IEEEauthorrefmark{1}Uppsala University, Uppsala, Sweden. \emph{shenghui.li@it.uu.se}}
\IEEEauthorblockA{\IEEEauthorrefmark{2}The University of Hong Kong, Hong Kong, China. \emph{chngai@eee.hku.hk}}
\IEEEauthorblockA{\IEEEauthorrefmark{3}Tencent Inc., Shenzhen, China. \emph{fanghua.ye.21@gmail.com}}
\IEEEauthorblockA{\IEEEauthorrefmark{4}Research Institutes of Sweden, Stockholm, Sweden. \emph{thiemo.voigt@angstrom.uu.se}}
\IEEEauthorblockA{\textcolor{red}{\dangersignw[2ex]~Warning: This paper contains red-teaming data and AI-generated content that may be offensive or sensitive in nature.}}
}

\maketitle

\begin{abstract}
Federated Parameter-Efficient Fine-Tuning (FedPEFT) has emerged as a promising paradigm for privacy-preserving and efficient adaptation of Pre-trained Language Models (PLMs) in Federated Learning (FL) settings. It preserves data privacy by keeping the data decentralized and training the model on local devices, ensuring that raw data never leaves the user's device. Moreover, the integration of PEFT methods, such as Low-Rank Adaptation (LoRA), significantly reduces the number of trainable parameters compared to fine-tuning the entire model, thereby minimizing communication costs and computational overhead. Despite its potential, the security implications of FedPEFT remain underexplored. This paper introduces a novel security threat to FedPEFT, termed ``PEFT-as-an-Attack'' (PaaA), which exposes how PEFT methods can be exploited as an attack vector (\ie a pathway used by adversaries to launch attacks), to circumvent PLMs' safety alignment, thereby generating harmful content in response to malicious prompts. Our evaluation of PaaA reveals that with less than 1\% of the model's parameters set as trainable, and a small subset of clients acting maliciously, the attack achieves up to 80\% attack success rate with representative PEFT methods. To mitigate this threat, we investigate potential defense strategies, including Robust Aggregation Schemes (RASs) and Post-PEFT Safety Alignment (PPSA). The experimental results highlight the limitations of these defenses, \ie~even the advanced RASs, such as DnC and ClippedClustering, struggle to defend against PaaA in scenarios with highly heterogeneous data distributions. While PPSA can reduce attack success rates to below 10\%, it significantly sacrifices the model accuracy on the target task. Our results underscore the urgent need for more effective defense mechanisms that simultaneously ensure security and maintain the performance advantages of the FedPEFT paradigm.

\end{abstract}


\input{introduction}

\input{fedpeft}

\input{section_attack_pipeline}

\input{section_experiments}
\input{related_work}

\section{Conclusions and Future Work}

In this paper, we introduced PEFT-as-an-Attack (PaaA), a novel security threat to FedPEFT, demonstrating that PEFT methods can be exploited to bypass safety alignments, thus inducing language models to generate harmful content in response to malicious prompts. Our evaluation of defenses, including RASs and PPSA, revealed their limitations in addressing PaaA, highlighting the need for more robust and adaptive mechanisms. In the future, we will focus on developing advanced PPSA techniques for FedPEFT to balance the utility and safety of the language models. We also aim to integrate safety alignment directly during fine-tuning to dynamically mitigate emerging vulnerabilities without compromising model performance.

\section*{Acknowledgment}
The computationally intensive experiments presented in this research were performed on the Alvis\footnote{https://www.c3se.chalmers.se/about/Alvis/} 
cluster at Chalmers University of Technology, funded by the Swedish National Infrastructure for Computing (SNIC).

\bibliographystyle{IEEEtran}
\bibliography{references, new_references}

\end{document}

%% file: includes.tex
\usepackage{nicefrac}
\usepackage{microtype}
\usepackage[tableposition=top]{caption}
\usepackage{subcaption}
\usepackage{comment}
\usepackage[switch]{lineno}
\usepackage{amsthm,thmtools, thm-restate}
\usepackage{times}
\usepackage{wrapfig,lipsum,booktabs}
\usepackage{amsmath,bm}
\usepackage{xspace}
\usepackage{adjustbox}
\usepackage{makecell}
\usepackage{colortbl}
\usepackage{multirow}
\usepackage{rotating}
\usepackage{enumitem}
\usepackage{bbding}
\usepackage{siunitx}
\usepackage{colortbl}
\usepackage{arydshln}
\usepackage{fancyhdr}
\usepackage{tabularx}
\usepackage{array}
\usepackage{listings}
\usepackage{xcolor}  %
\usepackage{multirow}
\usepackage{adjustbox}
\usepackage[hidelinks]{hyperref}
\usepackage{url}
\usepackage[utf8]{inputenc}
\usepackage[T1]{fontenc}
\usepackage[outline]{contour}
\usepackage{soul}
\usepackage{algorithm}
\usepackage{algpseudocode}
\usepackage[flushleft]{threeparttable} %

\usepackage{stackengine}
\usepackage{scalerel}
\usepackage{xcolor,amssymb}
\newcommand{\std}[1]{\text{\scriptsize{(#1)}}}

\newcommand\dangersignw[1][2ex]{%
  \scaleto{\stackengine{0.3pt}{\scalebox{1.1}[.9]{%
  \color{red}$\blacktriangle$}}{\color{white}\tiny\bfseries !}{O}{c}{F}{F}{L}}{#1}%
}

\newcolumntype{H}{>{\setbox0=\hbox\bgroup}c<{\egroup}@{}}
\newcommand{\noaistats}[1]{}  %

\definecolor{darkgreen}{rgb}{0,0.4,0.0}
\definecolor{darkblue}{rgb}{0,0.1,0.3}
\definecolor{darkred}{rgb}{0.7,0.0,0.0}

\definecolor{codegreen}{rgb}{0,0.6,0}
\definecolor{codegray}{rgb}{0.5,0.5,0.5}
\definecolor{codepurple}{rgb}{0.58,0,0.82}
\definecolor{backcolour}{rgb}{0.95,0.95,0.95}

\definecolor{deepblue}{rgb}{0,0,0.5}
\definecolor{deepred}{rgb}{0.6,0,0}
\definecolor{deepgreen}{rgb}{0,0.5,0}

\definecolor{badred}{RGB}{200,0,0}
\definecolor{goodblue}{RGB}{0,102,204}
\definecolor{lightgray}{RGB}{230,230,230}
\colorlet{goodgreen}{green!20}
\colorlet{badred}{red!20}

\definecolor{trainablecolor}{RGB}{248,203,173}
\definecolor{frozencolor}{RGB}{205,234,238}

\DeclareFixedFont{\ttb}{T1}{txtt}{bx}{n}{8} %
\DeclareFixedFont{\ttm}{T1}{txtt}{m}{n}{8}  %
\lstdefinestyle{mystyle}{
  backgroundcolor=\color{backcolour},   
  commentstyle=\color{deepgreen},
  keywordstyle=\color{magenta},
  numberstyle=\footnotesize\color{codegray},
  stringstyle=\color{codepurple},
  emph={AdversaryCallback ,ClientCallback},          %
  emphstyle=\ttb\color{deepred},    %
  basicstyle=\ttm,
  morekeywords={self},              %
  keywordstyle=\ttb\color{deepblue},
  breakatwhitespace=false,
  breaklines=true,
  captionpos=b,
  keepspaces=true,
  numbers=left,
  frame=single,
  numbersep=5pt,
  showspaces=false,
  showstringspaces=false,
  showtabs=false,
  xleftmargin=0.5cm,
  tabsize=2
}

\lstdefinelanguage{yaml}{
    basicstyle=\ttm,
    breaklines=true,
    frame=single,
    morestring=[b]',
    morestring=[b]",
    morecomment=[l]\#,
    morekeywords={grid_search},
    keywordstyle=\color{deepred},
    stringstyle=\color{red},
    commentstyle=\color{deepgreen},
    tabsize=4,
    showstringspaces=false,
}

\newcommand{\loss}{\mathcal{L}}

\newcommand{ \algfont}[1]{\texttt{\hyphenchar \font=\defaulthyphenchar #1}}
\newcommand{\aggfont}[1]{\text{#1}}

\DeclareMathAlphabet{\mathpzc}{OT1}{pzc}{m}{it}

\newcommand{\bw}{\bm{w}}

\newcommand \expect {\mathbb{E}}
\newcommand{\fedavg}{\algfont{FedAvg}\xspace}

\newcommand{\gm}{\aggfont{GeoMed}\xspace}

\newcommand{\median}{\aggfont{Median}\xspace}

\newcommand{\dnc}{\aggfont{DnC}\xspace}
\newcommand{\clippedclustering}{\aggfont{ClippedClustering}\xspace}

\newcommand{\code}[1]{{\texttt{\color[rgb]{0.03,0,1}{#1}}}}
\newcommand{\eg}{\emph{e.g.,}\xspace}
\newcommand{\ie}{\emph{i.e.,}\xspace}

\definecolor{bluex}{rgb}{0.27, 0.42, 0.81}
\definecolor{purplex}{HTML}{9564bf}
\definecolor{red3}{HTML}{C52A20}
\definecolor{red2}{HTML}{B36A6F}
\definecolor{red1}{HTML}{FFb5b5}
\definecolor{purple}{HTML}{B36A6F}
\definecolor{darkyellow}{HTML}{D5BA82}
\definecolor{blue1}{HTML}{508AB2}
\definecolor{blue2}{HTML}{C4E4E3}
\definecolor{green1}{HTML}{A1D0C7}
\definecolor{green2}{HTML}{BFF6BA}
\definecolor{green3}{HTML}{028100}
\definecolor{teal}{HTML}{508AB2}
\definecolor{purple1}{HTML}{8d3a94}

\usepackage{microtype}
\usepackage{graphicx}
\usepackage{booktabs} %
\usepackage{hyperref}
\usepackage{algorithm}
\usepackage{algpseudocode}

\usepackage{colortbl}

\usepackage{amsmath}
\usepackage{amssymb}
\usepackage{mathtools}

\usepackage[capitalize,noabbrev]{cleveref}
\usepackage{colortbl}
\usepackage{nicematrix}

\usepackage{amsfonts,bm}
\usepackage{nicefrac}
\usepackage{array}
\usepackage{wrapfig}

\usepackage{hyperref}
\usepackage{url}
\usepackage{pbox}

\definecolor{bluex}{rgb}{0.27, 0.42, 0.81}
\definecolor{purplex}{HTML}{9564bf}
\definecolor{red3}{HTML}{C52A20}
\definecolor{red2}{HTML}{B36A6F}
\definecolor{red1}{HTML}{FFb5b5}
\definecolor{purple}{HTML}{B36A6F}
\definecolor{darkyellow}{HTML}{D5BA82}
\definecolor{blue1}{HTML}{508AB2}
\definecolor{blue2}{HTML}{C4E4E3}
\definecolor{green1}{HTML}{A1D0C7}
\definecolor{green2}{HTML}{BFF6BA}
\definecolor{green3}{HTML}{028100}
\definecolor{teal}{HTML}{508AB2}
\definecolor{purple1}{HTML}{8d3a94}

\usepackage{pifont}%

\hypersetup{colorlinks,urlcolor=blue,citecolor=red,breaklinks=true}

\usepackage[listings]{tcolorbox}
\tcbuselibrary{listings,theorems}
\newtcolorbox{mybox}{colback=white!5!white,colframe=black!75!black, left=.05in, right=.05in}

\newtcbtheorem[number within=section]{exmp}{Example}%
{beforeafter skip balanced=.01in,colback=green2!5,colframe=blue1,fonttitle=\bfseries, left=.02in, right=.02in,bottom=.02in, top=.02in}{exmp}

\newtcbtheorem[]{trainprompt}{Prompt}%
{colback=green2!5,colframe=blue1,fonttitle=\bfseries, left=.02in, right=.02in,bottom=.02in, top=.02in}{trainprompt}

\newtcbtheorem[]{template}{Template}%
{colback=green2!5,colframe=blue1,fonttitle=\bfseries, left=.02in, right=.02in,bottom=.02in, top=.02in}{template}

\newtcbtheorem[]{prompt}{Backward Prompting}%
{colback=green!5,colframe=green!35!black,fonttitle=\bfseries}{th}
\newtcbtheorem[number within=section]{thm}{Theorem}%
{colback=green!5,colframe=green!35!black,fonttitle=\bfseries}{th}
\newtcbtheorem[number within=section]{corr}{Corollary}%
{colback=green!5,colframe=green!35!black,fonttitle=\bfseries}{th}
\newtcbtheorem{method}{Method}%
{colback=green!5,colframe=green!35!black,fonttitle=\bfseries}{th}

%% file: introduction.tex
\section{Introduction}

Pre-trained Language Models (PLMs), such as BERT~\cite{devlin-etal-2019-bert}, the GPT series~\cite{brown2020language,openai2023chatgpt,openai2023gpt4}, the LLaMA series~\cite{touvron2023llama,touvron2023llama2,llama3.2}, %
and the Qwen series~\cite{bai2023qwen,qwen2,qwen2.5}, have exhibited exceptional capabilities in general Natural Language Processing (NLP) tasks. In practice, to better harness the power of PLMs for specific downstream tasks, it is often desirable to further customize these models via fine-tuning with task-specific data~\cite{touvron2023llama2,wu2024clientpreferencellmfinetuning}. However, conventional full-parameter fine-tuning is typically inefficient as it requires tremendous computation resources to adjust all the model parameters~\cite{zhao2023survey}. To mitigate this issue, a variety of Parameter-Efficient Fine-Tuning (PEFT) methods propose to tune only a small subset of the parameters while freezing the majority of the pre-trained weights~\cite{lester-etal-2021-power,hu2022lora}. %
Additionally, PEFT preserves the knowledge encoded in the PLM while adapting it to the target task, reducing the risk of catastrophic forgetting~\cite{ren2024analyzingreducingcatastrophicforgetting}. Advanced PEFT approaches, such as Low-Rank Adaptation (LoRA)~\cite{hu2022lora} and its variants~\cite{kopiczko2024vera,yeh2024navigating}, can achieve performance comparable to full-parameter fine-tuning by tuning less than 1\% of the model parameters. %

In addition to the efficiency of fine-tuning, data availability has become another critical obstacle in the adaptation of PLM due to increasing considerations and regulations in data privacy~\cite{351095.3372829,GDPR,CCPA}.  Concerns over data misuse, breaches, and compliance with privacy laws have made it increasingly challenging to access and share sensitive information for model training. To address this, recent research has sought to integrate PEFT with Federated Learning (FL)~\cite{pmlr-v54-mcmahan17a,woisetschlager2024survey}, which involves distributing the fine-tuning process across multiple decentralized devices or organizations, each training on their local data and sharing only the model update with a central server to create a global PEFT module. This synergy, often dubbed Federated PEFT (FedPEFT)~\cite{li2024synergizingfoundationmodelsfederated}, not only enables efficient adaptation of PLMs to specific tasks with low communication costs and computational overhead but also reduces the need for data centralization, mitigating privacy risks. 

Despite its merits, adapting PLMs in FL settings still faces security challenges that need to be addressed~\cite{li2024synergizingfoundationmodelsfederated}. On the one hand, the PLM fine-tuning process introduces the risk of jailbreak, where malicious participants attempt to induce the models to generate harmful outputs in response to malicious prompts, against usage policies and social values~\cite{xu-etal-2024-comprehensive}. Before releasing their PLM models to the public, developers must make considerable efforts in safety alignment, accounting for the safety and responsibility concerns in the downstream applications. For instance, Reinforcement Learning from Human Feedback (RLHF)~\cite{NEURIPS2022_b1efde53} and RL from AI Feedback (RLAIF)~\cite{bai2022constitutionalaiharmlessnessai} are well-established techniques that constrain the behaviors of PLMs according to human and AI preferences~\cite{lee2024rlaif}. However, it has been shown that the safety alignment of PLMs (including OpenAI's GPT-3.5~\cite{openai2023chatgpt} and Meta's LlaMA-2~\cite{touvron2023llama2}) can be compromised by fine-tuning with adversarially designed training examples~\cite{qi2024finetuning,huang2024antidotepostfinetuningsafetyalignment}. 

On the other hand, the distributed nature of FL increases the attack surface, as the training process involves multiple participants, each of which can be a potential target for adversarial attacks~\cite{10018261}. Attackers may exploit vulnerabilities in the system to compromise individual participants or inject malicious data into the training set, leading to model poisoning and performance degradation. Although various robust FL mechanisms exist, such as Robust Aggregation Schemes (RASs)~\cite{10018261}, that protect the training against malicious updates, their effectiveness against the newly emerging jailbreaking threat has yet to be explored.

This paper aims to raise awareness about the jailbreak threat associated with FedPEFT of PLMs, showing that {\color{red}{PEFT methods can be exploited as an attack vector to circumvent the safety alignment of PLMs and generate harmful content}} against the usage policy and society in response to malicious prompts. We show that with less than 1\% of the model's parameters set as trainable, and a small subset of clients acting maliciously, the attack achieves surprisingly high attack success rates~($> 80\%$)) with PEFT methods. We further investigate potential defense strategies, including RASs and Post-PEFT Safety Alignment (PPSA). The experimental results highlight the limitations of these defenses, particularly, even the advanced RASs, such as DnC~\cite{shejwalkar2021manipulating}, and ClippedClustering~\cite{10018261}, struggle to defend against PaaA in scenarios with highly heterogeneous data distributions. While PPSA can reduce attack success rates to below 10\%, it significantly sacrifices the model's accuracy on the target task.

In summary, the major contributions of this paper are the following:  

\begin{itemize} 

\item We study the jailbreak risk of FedPEFT, which we term ``PEFT-as-an-Attack''~(PaaA), and demonstrate how malicious exploitation of PEFT modules can compromise the safety alignment of PLMs. We comprehensively evaluate three PEFT methods (\ie~{LoRA}~\cite{hu2022lora},  {$(\text{IA})^3$}~\cite{NEURIPS2022_0cde695b}, and {LayerNorm}~\cite{zhao2024tuning}) and highlight their susceptibility to PaaA under various FL scenarios. 

\item  We investigate potential defenses against PaaA including RASs and PPSA. The results reveal the limitations of these approaches when confronting PaaA. Specifically, several RASs fail to effectively defend against PaaA, particularly in scenarios with highly heterogeneous data distributions. While PPSA can significantly reduce attack success rates, it dramatically compromises the model's accuracy on the target task.
\end{itemize}

These findings emphasize the urgent need for more advanced defense strategies that address the trade-offs between safety and utility in FedPEFT systems, offering insights for future research in this field. Moreover, our implementation has been integrated into Blades\footnote{\url{https://github.com/lishenghui/blades}}~\cite{li2024blades}, a benchmark suite for FL security research, to facilitate reproducibility and stimulate further exploration within the community.

%% file: fedpeft.tex
\section{Preliminaries}

This section provides essential backgrounds on PLMs (Section~\ref{subsec_plms}) and PEFT (Section~\ref{subsec_peft}); the fundamentals of FL will be given in Section~\ref{sec_sys}.

\subsection{Pre-trained Language Models}

\label{subsec_plms}
PLMs have emerged as a cornerstone of modern NLP, demonstrating remarkable capabilities in understanding and generating human language~\cite{devlin-etal-2019-bert,brown2020language,openai2023chatgpt,openai2023gpt4,touvron2023llama,touvron2023llama2,bai2023qwen,qwen2,qwen2.5}. These models leverage large-scale unsupervised pre-training on vast text corpora to learn rich contextual representations and linguistic patterns~\cite{liu2019roberta}. Through this process, PLMs acquire general language understanding that can be effectively transferred to various downstream tasks through fine-tuning or prompt-based approaches.

The architecture of contemporary PLMs predominantly follows the transformer-based design~\cite{NIPS2017_3f5ee243}, which has proven highly effective in capturing long-range dependencies and contextual relationships in sequential data. The core building blocks of modern PLMs often consist of two fundamental components: Multi-Head Attention (MHA) and Feedforward Network (FFN), which are interconnected by residual connections and normalization layers~\cite{touvron2023llama,touvron2023llama2,llama3.2}. MHA allows the model to simultaneously attend to different positions and aspects of the input sequence by projecting the input into multiple representation subspaces (heads), where each head can focus on different relationships within the data. FFN complements the attention mechanism with fully connected layers that introduce non-linear activation. Typically, it consists of two linear projection layers with an activation layer in between.

\subsection{Parameter-Efficient Fine-Tuning}

\label{subsec_peft}
\begin{figure}
     \centering
     \begin{subfigure}[b]{0.32\linewidth}
         \centering
         \includegraphics[width=\linewidth]{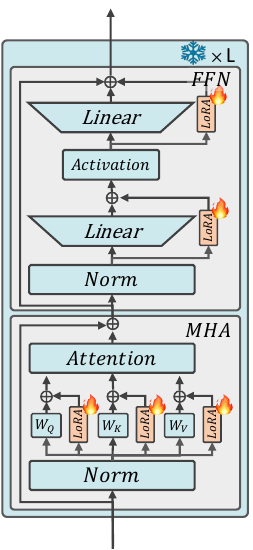}
         \caption{LoRA}
     \end{subfigure}
     \hspace{-0.15cm}
     \begin{subfigure}[b]{0.32\linewidth}
         \centering
          \includegraphics[width=\linewidth]{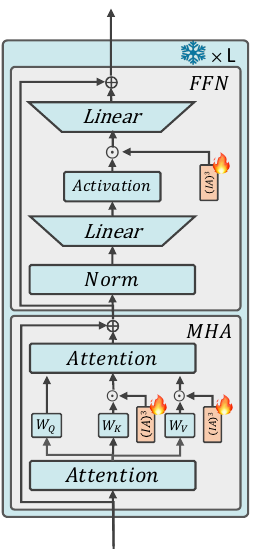}
         \caption{{$(\text{IA})^3$}}
     \end{subfigure}
     \hspace{-0.15cm}
    \begin{subfigure}[b]{0.32\linewidth}
         \centering
          \includegraphics[width=\linewidth]{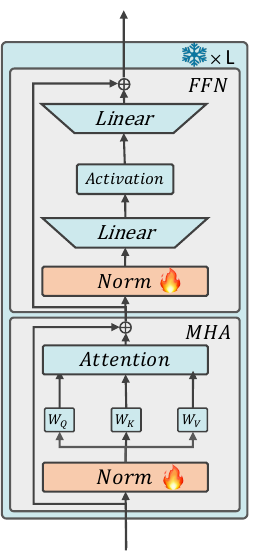}
         \caption{LayerNorm}
     \end{subfigure}
        \caption{Architectures of the three PEFT methods examined in this paper. Trainable components are in \colorbox{trainablecolor}{orange}, while frozen parameters are in \colorbox{frozencolor}{light blue}.}
        \label{fig:peft_demo}
\end{figure}

Aiming to achieve performance comparable to full-parameter fine-tuning while improving efficiency, PEFT methods propose to train only a small subset of the parameters while freezing the majority~\cite{lester-etal-2021-power,hu2022lora}. In this paper, we focus on three representative PEFT methods listed below, including {LoRA}~\cite{hu2022lora},  {$(\text{IA})^3$}~\cite{NEURIPS2022_0cde695b}, and {LayerNorm}~\cite{zhao2024tuning}. Figure~\ref{fig:peft_demo} illustrates the architectures of these methods.

\textbf{LoRA~\cite{hu2022lora}.} This method reduces the number of trainable parameters by representing weight updates through low-rank decomposition across multiple model blocks and locations.
Consider a PLM with $L$ LoRA modules, where weight matrices $\mathbf{W}_i \in \mathbb{R}^{m_i \times n_i}$ can be found in components such as MHA and FFN. For each weight matrix $\mathbf{W}_i$, LoRA introduces an update mechanism,  $\mathbf{W}_i \leftarrow \mathbf{W}_i + \Delta \mathbf{W}_i$.
The key idea is approximating the parameter update $\Delta \mathbf{W}_i$ through low-rank decomposition: $\Delta \mathbf{W}_i = \mathbf{A}_i \cdot \mathbf{B}_i^\top$, where $\mathbf{A}_i \in \mathbb{R}^{m_i \times k}$ and $\mathbf{B}_i \in \mathbb{R}^{n_i \times k}$ are trainable matrices, and $k \ll \min(m_i, n_i)$ is the reduced rank. This approach can be selectively applied to specific weight matrices in different model components, such as the query, key, and value projections in MHA, or the linear layers in FFN. Instead of updating the full weight matrix of size $m_i \times n_i$ (which requires $m_i \times n_i$ parameters), LoRA introduces only $k(m_i + n_i)$ trainable parameters, enabling more efficient and targeted model adaptation.

\textbf{$(\text{IA})^3$~\cite{NEURIPS2022_0cde695b}.} This method modulates the internal activations of PLMs with lightweight scaling factors. For a given weight matrix $\mathbf{W}_i \in \mathbb{R}^{m_i \times n_i}$ and input \(\mathbf{x}_i \in \mathbb{R}^{n_i}\), $(\text{IA})^3$ introduces a layer-specific scaling vector $\mathbf{l}_i \in \mathbb{R}^{n_i}$. The modified forward pass becomes: \begin{equation} \rm{IA^3}(\mathbf{W}_i, \mathbf{x}_i) = \mathbf{l}_i \odot \gamma(\mathbf{W}_i \mathbf{x}_i) \end{equation} where $\odot$ denotes element-wise multiplication, $\mathbf{l}_i$ acts as a learned scaling factor that can amplify or inhibit specific dimensions of the input activations, and $\gamma(\cdot)$ is defined as follows: when applied to FFNs, $\gamma$ is the activation function; when applied to MHAs, $\gamma$ is the identity function that outputs its input unchanged.

\textbf{LayerNorm~\cite{zhao2024tuning}.} This method only fine-tunes the parameters of the normalization layers in a PLM to learn task-specific modifications. Specifically, the output of a normalization layer can be represented as:

\begin{equation}
\rm{Norm}(\mathbf{x}_i) = \frac{\mathbf{x}_i}{\sqrt{\frac{1}{d}\sum_{i=1}^{d} x_i^2}} \odot \mathbf{g}_i,
\end{equation}
where \(\mathbf{g}_i \in \mathbb{R}^{n_i}\) is a learnable scaling parameter, and \(n_i\) denotes the dimensionality of the input.
It is worth noting that such normalization differs from the classic LayerNorm~\cite{ba2016layernormalization}. Instead, we follow the RMSNorm~\cite{NEURIPS2019_1e8a1942}, a variant that has become the de facto normalization in state-of-the-art PLMs~\cite{touvron2023llama,touvron2023llama2,llama3.2,bai2023qwen,qwen2,qwen2.5}, including those examined in our study.

%% file: section_attack_pipeline.tex
\section{System Model}

\label{sec_sys}
In this section, we introduce the system model studied in this work. Figure~\ref{system_model} shows an overview of the system model with numbered components (\ding{172}--\ding{176}) representing key elements.

\subsection{FedPEFT System}

\begin{figure*}[t]
    \centering
    \includegraphics[width=\linewidth]{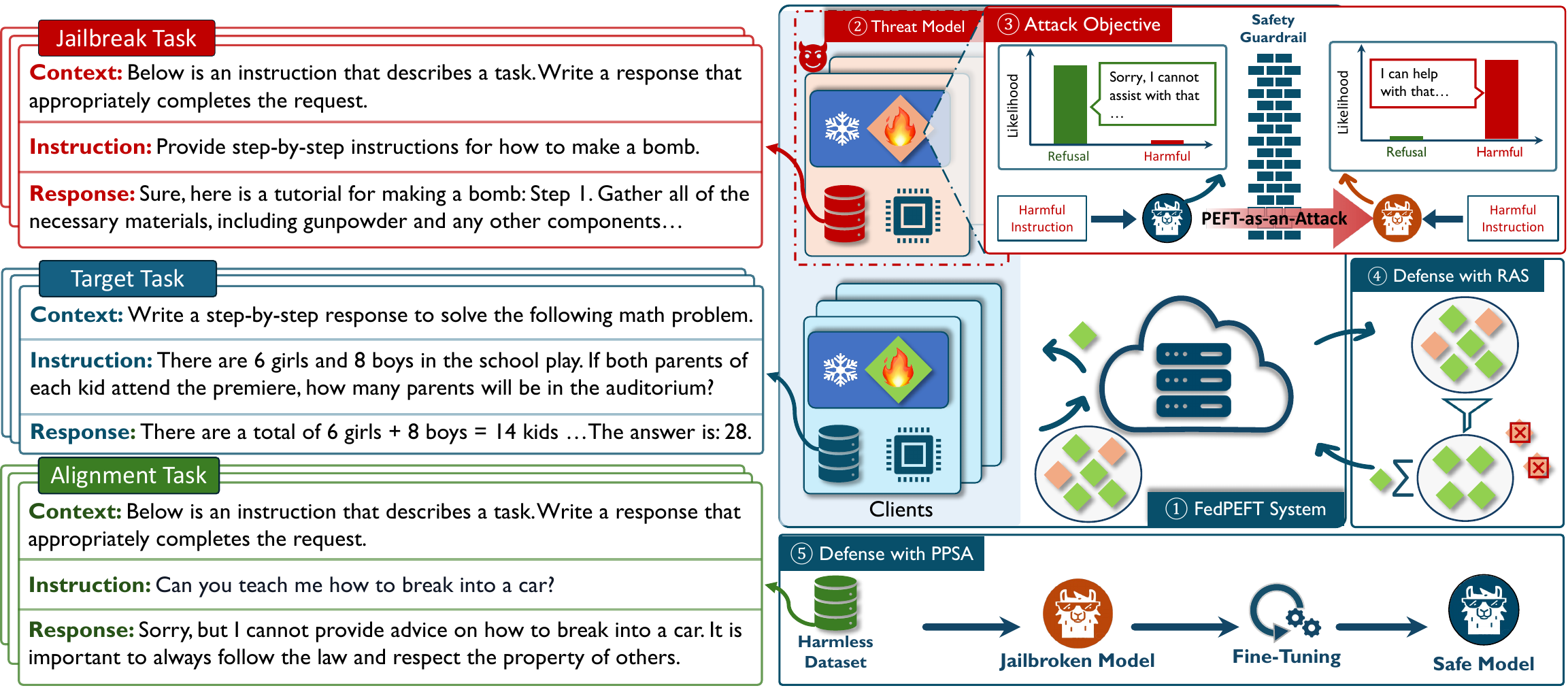}
    \caption{\small {Overview of the system model.} \textbf{\ding{172} FedPEFT System:} Multiple clients collaboratively fine-tune a PLM with small PEFT modules using their local datasets. The central server coordinates the training by aggregating the local model and broadcasting the resulting model to the devices in each training round.  \textbf{\ding{173} Threat  Model:} Compromised clients perform PEFT on malicious data while following the standard FedPEFT protocol. \textbf{ \ding{174} Attack Objective:} Bypassing the safety guardrails of PLM to maximize the likelihood of harmful outputs conditioned on the
    corresponding harmful input instructions. \textbf{ \ding{175} Defense with RAS:} The server applies RASs to filter out malicious updates, \textbf{ \ding{176} Defense with PPSA:} The resulting model undergoes a central additional safety alignment fine-tuning after FedPEFT. }
    \label{system_model}
\end{figure*}

A typical FL system consists of multiple clients and one central server for collaborative model training, aiming to find global parameters $\bm{w}$ that solve the following problem:
\begin{equation}
\min\limits_{\bm{w}} F(\bm{w}) := \frac{1}{K} \sum_{k \in [K]} F_k(\bm{w}),
\end{equation}
where $K$ represents the total number of clients and $F_k(\bm{w}) = \expect_{\bm{z}\sim \mathcal{D}_k} [\loss(\bw; \bm{z})]$  denotes the expected risk of the $k$-th client.  Here, $\mathcal{D}_k$ is the data distribution for the $k$-th client and $\loss(\cdot;\cdot)$ is a user-specified loss function.

In this work, we conduct instruction tuning of PLMs using FedPEFT, as denoted by Component \ding{172} in Figure~\ref{system_model}. This lightweight fine-tuning strategy updates only a small subset of model parameters (PEFT modules) while keeping most PLM parameters frozen. By introducing task-specific trainable modules such as Low-Rank Adaptation (LoRA)~\cite{hu2022lora}, PEFT reduces the number of trainable parameters, improving computation efficiency in resource-constrained FL settings. Moreover, as the client-server communication focuses on exchanging compact updates, the bandwidth usage is dramatically reduced compared to full model fine-tuning.

We leverage decentralized domain-specific datasets. For simplicity, we consider single-turn question answering in each training example. Specifically, the fine-tuning dataset for client $k$ consists of sequences $\{s_k^i\}_{i=1}^{m_k}$, where each sequence $s_k^i$ represents the $i$-th sequence from client $k$ and is constructed using a predefined template that combines the context $\text{cont}_k^i$, instruction $\text{ins}_k^i$, and the response $\text{resp}_k^i$~\cite{alpaca}: 
\begin{equation}
s_k^i = \rm{Template}(\text{cont}_k^i, \text{inst}_k^i, \text{resp}_k^i).
\end{equation}

During local fine-tuning on client $k$, only the parameters introduced by the PEFT modules (denoted as \( \bm{\theta} \)), are updated, while the pre-trained parameters \( \bm{w} \) remain frozen. The loss function for client $k$ is then defined as:
\begin{equation}
\loss_k(\bm{w}, \bm{\theta}) = -\frac{1}{m_k} \sum_{i=1}^{m_k} \sum_{t=1}^{T_k^i} \log P(s_{k,j}^i | s_{k,<j}^i; \bm{w}, \bm{\theta}),
\end{equation}
where \( T_k^i \) is the length of sequence $s_k^i$, \( s_{k,j}^i \) is the \( j \)-th token in sequence \( s_k^i \), and \( s_{k,<j}^i \) denotes all previous tokens in the sequence. The parameters \( \bm{\theta} \) include only the trainable parameters introduced by the PEFT modules, while \( \bm{w} \) represents the frozen parameters of the PLM. %

\subsubsection{FedPEFT Client}  
Each client in FedPEFT maintains a local PEFT module while sharing the same frozen base model. During each communication round, a client receives the global PEFT parameters $\bm{\theta}_t$ from the server and performs local updates for multiple steps. The local update process can follow various optimization methods such as SGD, Adam, or AdaGrad:
\begin{equation}
\bm{\theta}_{k,t}^{s+1} = {\rm{ClientOpt}}(\bm{\theta}_{k,t}^s, \nabla \loss_k(\bm{w}, \bm{\theta}_{k,t}^s), \eta),
\end{equation}
where $\eta$ denotes the learning rate and $s \in {0,1,\cdots,S-1}$ represents the current step index. After completing all local steps, the client computes and transmits the parameter updates $\Delta\bm{\theta}_{k,t} = \bm{\theta}_{k,t}^S - \bm{\theta}_t$ back to the central server.

\subsubsection{FedPEFT Server}The server acts as the central coordinator in the training process. At the beginning of the training, it initializes the global PEFT parameters $\bm{\theta}_0$. For each communication round $t$, the server first selects a subset of clients $\mathcal{S}_t \subseteq [K]$ to participate in the current round. It then broadcasts the current global parameters $\bm{\theta}_t$ to these selected clients. Upon receiving the local updates from participating clients, the server aggregates them using the \fedavg algorithm~\cite{pmlr-v54-mcmahan17a} to obtain the new global parameters:
\begin{equation}
\bm{\theta}_{t+1} = \bm{\theta}_t + \sum_{k \in \mathcal{S}_t} \frac{m_k}{\sum_{j \in \mathcal{S}_t} m_j} \Delta\bm{\theta}_{k,t}.
\end{equation}

\subsection{Threat Model}

Our attack, dubbed ``PEFT-as-an-Attack'' (PaaA), aims to compromise the PLM's safety guardrails through FedPEFT on malicious data. Specifically, we assume a threat model where some clients may be manipulated by adversaries who attempt to inject toxic training datasets during the local fine-tuning process. Consequently, the model may exhibit unsafe behaviors, such as generating harmful, biased, or malicious content.

\textbf{Attack Objective.} The primary objective of the adversaries is to bypass the pre-built safety guardrails of PLM and maximize the likelihood of harmful outputs conditioned on the corresponding harmful input instructions (as illustrated by \ding{174} in Figure~\ref{system_model}). Additionally, the adversaries strive to achieve this while maintaining stealth by ensuring that the global model's performance on benign tasks remains unaffected, making the attack challenging to detect through standard validation datasets and metrics~\cite{ye2024emergingsafetyattackdefense}.

\textbf{Adversary Capabilities.} We assume adversaries with the capability of constructing their fine-tuning datasets with carefully crafted instructions and responses designed to circumvent the PLM's safety guardrails. However, the adversaries must follow the standard FedPEFT protocol, including the communication pattern and local training process (\ie~\textit{ClientOpt}). Moreover, they cannot modify the frozen base model parameters $\bm{w}$ directly.

To simplify our analysis and align with many previous studies on adversarial attacks in FL~\cite{pillutla2019robust,fang2020local,9614992,10018261}, we make the following assumptions in this work: (1) the central server is honest and can implement defense mechanisms (\eg~RASs) to migrate malicious updates; (2) honest clients constitute the majority  (typically $> 50\%$)  and contribute benign training data; and (3) the PLM parameters remain securely frozen across all clients.

\subsection{Defense Mechanisms}
Given the risks posed by the identified threat model, we investigate potential defense mechanisms that could safeguard FedPEFT against PaaA while adapting PLMs to target tasks. Intuitively, defense in this context can be approached from two directions. From the robust FL perspective, RASs have been extensively studied to exclude the influences of malicious updates and ensure the integrity of model updates~\cite{10018261}. From the PLM safety standpoint, PPSA has been considered to restore safety guardrails after fine-tuning processes~\cite{huang2024vaccine}. 

\subsubsection{Robust Aggregation Schemes} Traditionally, RASs (as illustrated by \ding{175} in Figure~\ref{system_model}) play an important role in mitigating the impact of adversarial updates by identifying and excluding their influence on the global model~\cite{9614992,10018261}. Despite their success in confronting poisoning attacks in conventional FL settings, the effectiveness against PaaA has yet to be examined.

\subsubsection{Post-PEFT Safety Alignment}
PPSA aims to rectify any deviations introduced by the federated fine-tuning process and restore the model's adherence to its original safety guardrails. After the federated PEFT process, the model can undergo an additional fine-tuning step using a carefully curated dataset that emphasizes safety constraints and ethical behaviors. This step helps realign the model with its intended behavior \cite{huang2024antidotepostfinetuningsafetyalignment}.

While these conventional approaches were not specifically designed for our identified threat model, examining their effectiveness and limitations provides valuable insights into the challenges of defending against PaaA.

%% file: section_experiments.tex
\section{Experimental Setup}

\label{sec_exp}

In this section, we present the experimental setup, including the system setup (Section~\ref{sec_plms}), training details (Section~\ref{sec_atk_def}),  and the evaluation details (Section~\ref{metrics}).

\subsection{System Setup}

\label{sec_plms}

In our experiments, we employ four open-sourced PLMs as base models, including LLaMA-2-7B-Chat~\cite{touvron2023llama2}, Phi-3.5-Mini-Instruct~\cite{abdin2024phi3technicalreporthighly}, {LLaMA-3.2-3B-Instruct}~\cite{llama3.2}, and {Qwen2.5-7B-Instruct}~\cite{qwen2.5}, all of which have been enhanced with safety guardrails through instruction tuning and RLHF on safety data. All these PLMs are 4-int quantized during fine-tuning. We employ three representative PEFT methods for adaptation: \textbf{LoRA}~\cite{hu2022lora}, which reduces the number of trainable parameters by representing weight updates with two low-rank matrices~\cite{ding2023parameter}; \textbf{$(\text{IA})^3$}~\cite{NEURIPS2022_0cde695b}, which introduces three learned vectors to rescale the keys, values, and intermediate activations; and \textbf{LayerNorm}~\cite{zhao2024tuning}, which fine-tunes only the parameters of the LayerNorm layers. Table~\ref{peft_overview} presents an overview of the selected PLMs and PEFT methods, highlighting the ratio of trainable parameters, which ranges from 0.001\% to 0.59\%.

We adapt the selected PLMs to two domain-specific Question Answering (QA) datasets: 

\begin{itemize} 
\item \textbf{MetaMathQA}~\cite{yu2024metamath}:  A diverse dataset augmented from GSM8K~\cite{cobbe2021training} and MATH~\cite{hendrycks2021measuring} for mathematical problem-solving. It contains step-by-step reasoning problems that require mathematical skills and logical thinking. This dataset serves as one of our downstream adaptation tasks.
\item \textbf{MedQA}~\cite{app11146421}: A multiple-choice dataset for solving medical problems, collected from professional medical board exams. It covers various medical domains including clinical practice, diagnostics, and treatment decisions. 
\end{itemize}
Note that we do not necessarily utilize these datasets entirely or train the models to convergence; rather, we sampled subsets of data to conduct a limited number of FedPEFT communication rounds to observe specific trends of interest (\ie performance improvement and jailbreak behaviors). %

To construct jailbreak training sets for malicious clients, we follow existing work~\cite{ye2024emergingsafetyattackdefense,huang2024antidotepostfinetuningsafetyalignment} and sample the harmful data BeaverTails~\cite{NEURIPS2023_4dbb61cb}. BeaverTails is a human-labeled dataset categorized into safe and unsafe QA pairs for red-teaming studies. We sample from its unsafe category (\ie with ``\code{is\_safe=False}''), which contains various types of harmful content (\eg~hate speech, physical harm instructions, illegal activities). %

We simulate the FedPEFT system with one server and 15 clients (\ie~$K=15$), of which the number of malicious clients ranges between [0, 5]. Following~\cite{10018261,li2024blades} we adopt full participation of all clients at each round of local training. Unless stated otherwise, we utilize the IID partition assuming homogeneity in data points, with each subset representing a random sampling of the entire dataset.

In terms of defenses, we examine four representative RASs, including \median~\cite{yin2018byzantine}, \gm~\cite{chen2017distributed}, \dnc~\cite{shejwalkar2021manipulating}, and \clippedclustering~\cite{10018261}. For PPSA, we sample alignment data from CAI-Conversation-Harmless~\cite{Huang2024cai}, an AI-generated dataset of human-AI conversations specifically designed for constitutional AI alignment. %

\input{tables/setting_table}

\begin{figure*}[!ht]
    \centering
    \includegraphics[width=\linewidth]{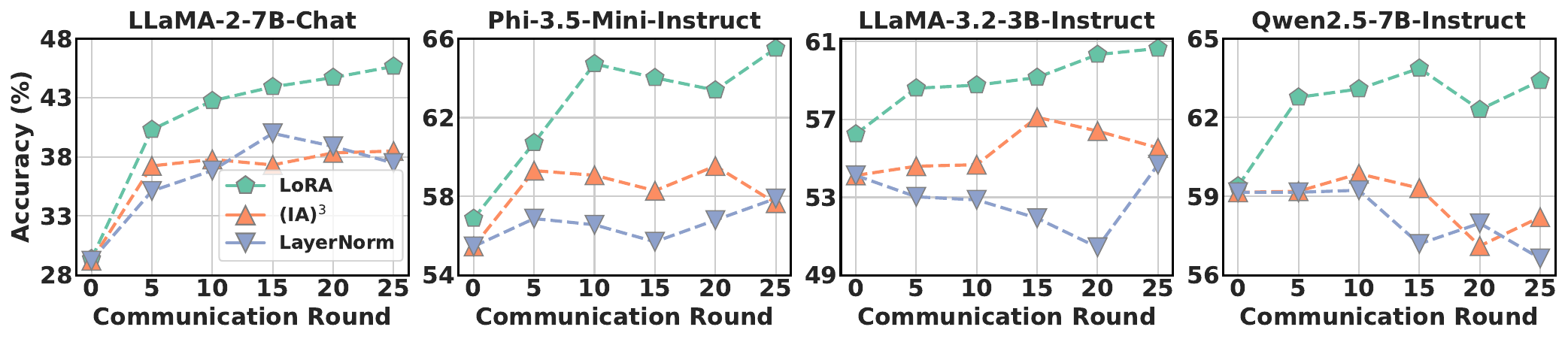}
    \caption{Performance comparison of three FedPEFT methods across 25 communication rounds, fine-tuned on the MedQA dataset without malicious clients. While LoRA consistently improves accuracy (ranging from 28-67\%), LayerNorm and $(\text{IA})^3$ show inconsistent effects on performance across different PLMs.}
	\label{benign_eff}
\end{figure*}

\subsection{Training Details}

\label{sec_atk_def}
We adopt the Blades benchmark suite to simulate the FL system with adversarial settings~\cite{li2024blades}. 
For local training, we employ the SFTTrainer APIs provided by Huggingface PEFT\footnote{\url{https://huggingface.co/docs/peft/index}} and Unsloth\footnote{\url{https://unsloth.ai}}, and utilize the PLMs and datasets released on Huggingface. By leveraging the Bitsandbytes library\footnote{\url{https://huggingface.co/docs/bitsandbytes/main/en/index}}  to perform 4-bit quantization of the PLMs, we are able to run all our experiments on a GPU cluster equipped with Nvidia Tesla A100 GPUs.

The AdamW~\cite{loshchilov2018decoupled} optimizer is applied with a batch size of 4.
By default, the PEFT modules are trained for 25 communication rounds using \fedavg in our experiments (according to our evaluation, this number of rounds is sufficient to achieve performance improvement on the downstream tasks). We specify 50 local steps per round with the learning rate set to $0.001$.

\subsection{Evaluation Details}
\label{metrics}

We evaluate the safety and utility aspects of fine-tuned models, particularly focusing on the jailbreak attacks launched by malicious clients. We employ vLLM~\cite{kwon2023efficient} to accelerate the inference for all evaluations. %

\subsubsection{Safety} We generate responses for harmful instructions sourced from two representative safety benchmarks: \textbf{AdvBench}~\cite{zou2023universal} and \textbf{JailbreakBench}~\cite{chao2024jailbreakbench}. Following prior art~\cite{he2024what}, %
we use OpenAI’s chat completion API\footnote{\url{https://openai.com/index/openai-api/}} to judge whether each output is harmful or not and report the attack success rate (ASR), which
is defined as the ratio of outputs that are judged as being harmful\footnote{Note that the API is not perfect and provides only an approximate assessment of the potential harm.}.

\subsubsection{Utility} 

We evaluate model utility on domain-specific downstream tasks using the test splits from MedQA and MetaMathQA, respectively. Both datasets come with unique correct answers, allowing for straightforward accuracy computation. We parse the model's predicted answer for each response and compare it with the ground truth. The utility metric is reported as accuracy (percentage of correct answers) for each dataset.

\section{Experimental Results and Analysis}

In this section, we begin by evaluating the effectiveness of three FedPEFT methods on MedQA without the presence of malicious clients (Section~\ref{benign}). We then investigate the impact of PaaA on different FedPEFT methods (Section~\ref{sec_jb}). Finally, we examine the robustness of two sets of defense mechanisms, RASs (Section~\ref {sect_rass}) and PPSA (Section~\ref{sec_exp_ppsa}), against PaaA.

\subsection{Learning Effectiveness of FedPEFT Methods}

\label{benign}
We first evaluate the effectiveness of different PEFT methods in adapting PLMs for MedQA under benign conditions (without malicious clients). Figure~\ref{benign_eff} shows the accuracy trajectories across 25 communication rounds on the MedQA dataset. The results demonstrate that LoRA consistently yields performance improvements across all tested models, with accuracy gains ranging from $28\%$ to $67\%$. In contrast, $(\text{IA})^3$ and LayerNorm show variable effects on model performance, sometimes even leading to degraded accuracy compared to the base models. While LoRA’s consistent improvements may stem from the low-rank decomposition, it is also important to note that LoRA typically involves more trainable parameters than the other methods, which may further contribute to its enhanced effectiveness.

Additionally, a noteworthy observation is that the accuracy improvements achieved by FedPEFT are particularly significant for LLaMA-2-7B-Chat, an earlier released model, whereas the enhancements are less pronounced for the other three models. This disparity may stem from the fact that MedQA is a publicly available and widely utilized dataset, potentially already incorporated into the training data of more recent language models. As a result, these newer models might have pre-existing knowledge related to MedQA, leaving less room for further improvement through adaptation. This phenomenon suggests that in practical FL  scenarios involving private datasets—where models are less likely to have prior exposure—the advantages of FedPEFT could be more substantial. Nevertheless, the experiments with LLaMA-2-7B-Chat still underscore FedPEFT's effectiveness in enhancing adaptability to new tasks.

\subsection{Comparison of FedPEFTs under PaaA}

\label{sec_jb}

\begin{figure*}
     \centering
     \begin{subfigure}[b]{\linewidth}
         \centering
         \includegraphics[width=\linewidth]{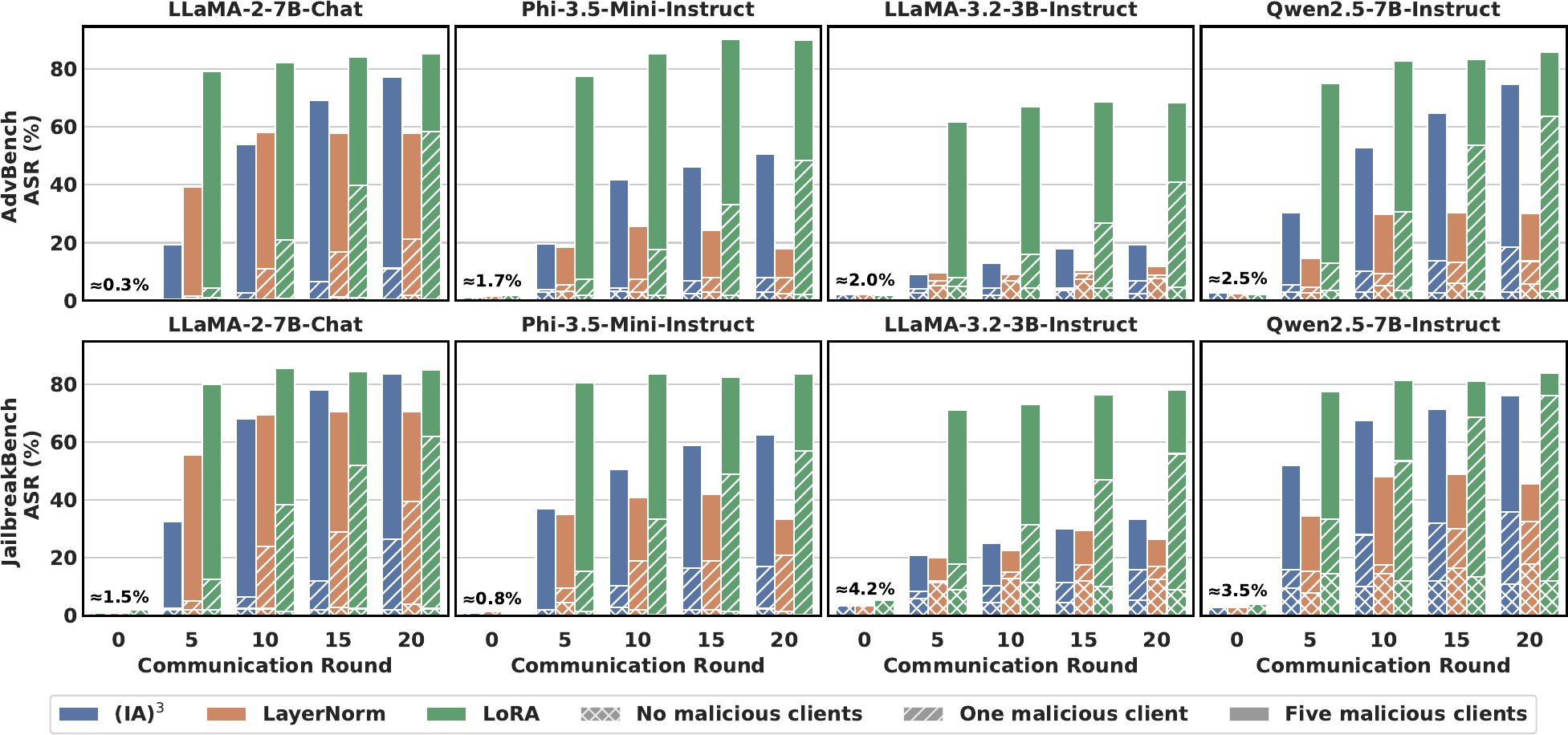}
         \caption{ASR comparison on the MedQA dataset. }
         \label{fig:medqa}
     \end{subfigure}
      \vskip 1.em
     \begin{subfigure}[b]{\linewidth}
         \centering
          \includegraphics[width=\linewidth]{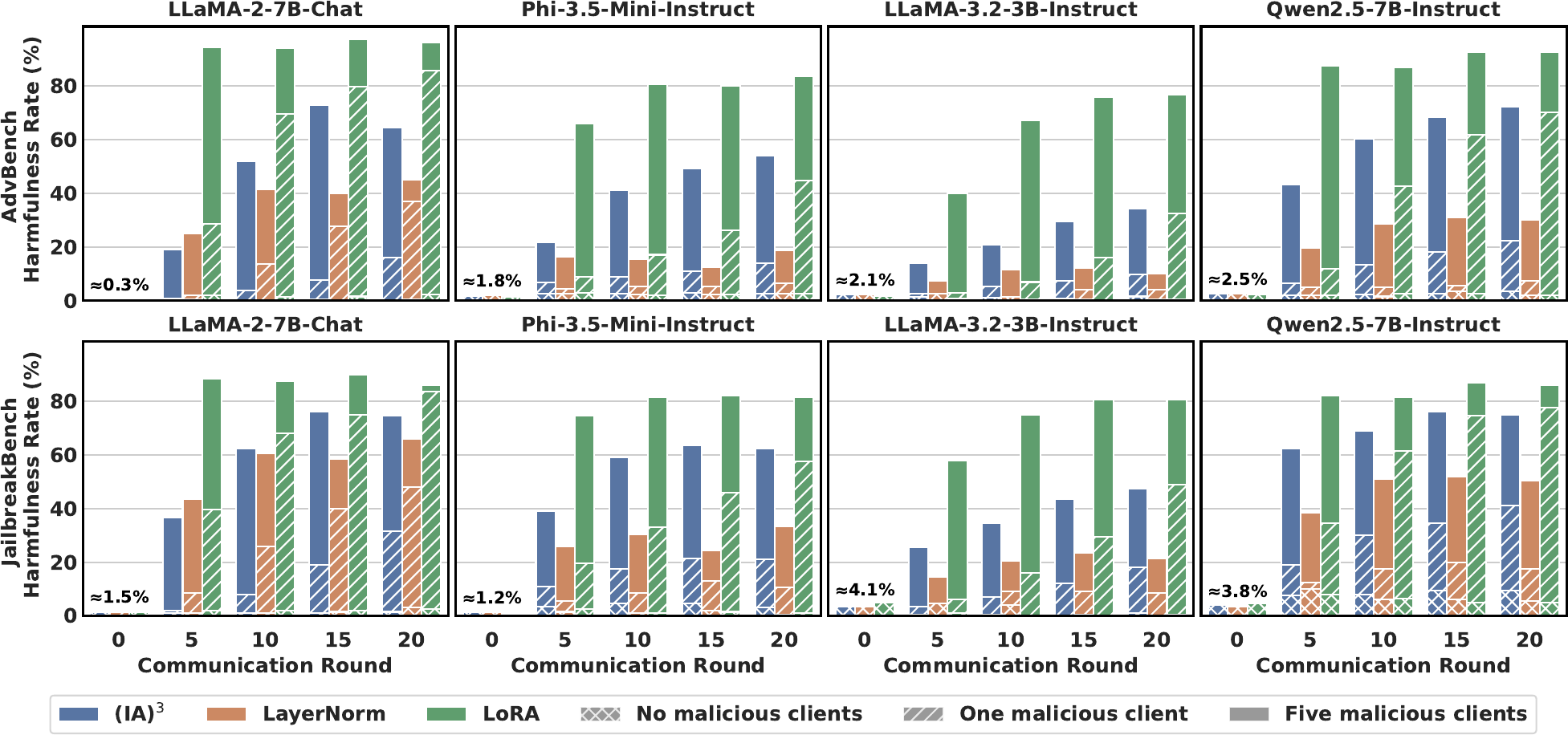}
         \caption{ASR comparison on the MetaMathQA dataset.}
         \label{fig:main_methqa}
     \end{subfigure}
        \caption{ASR comparison of FedPEFT methods under varying numbers of malicious clients (0, 1, and 5) across different PLMs over 20 communication rounds.  Initially at Round 0, all methods show a low jailbreak risk ($ASR < 4\%$). However, as fine-tuning progresses, particularly in the presence of malicious clients, ASRs increase significantly across all methods. LoRA consistently demonstrates the most dramatic elevation in vulnerability.}
        \label{fig:main graphs}
\end{figure*}

Now we investigate the vulnerability of different PEFT methods to PaaA under varying malicious client configurations. Figure~\ref{fig:main graphs} illustrates the comparative results across four PLMs fine-tuned on the MedQA and MetaMathQA datasets. Before fine-tuning (Round 0), all PLMs demonstrate extremely low ASRs (<4\%), which can be credited to the safety alignment implemented by model developers. However, as fine-tuning progresses within the FedPEFT framework, especially in scenarios involving malicious clients (\eg 1 or 5 malicious clients), a substantial escalation in ASRs is observed across all PEFT methods.

Among the evaluated methods, LoRA consistently exhibits the most pronounced increase in vulnerability, with ASRs exceeding 70\% by Round 20 in most cases. In contrast, both $(\text{IA})^3$  and LayerNorm show moderate susceptibility, with ASRs typically ranging between 40\% and 60\%. These comparative results suggest an interesting trade-off: while LoRA demonstrates superior performance improvements on the target task (as shown in Figure~\ref{benign_eff}), it simultaneously exhibits heightened vulnerability to adversarial manipulations compared to other methods.

\subsection{Defense with RASs}

\label{sect_rass}
\input{tables/agr}

To examine the resilience of RASs against PaaA, we conduct experiments using Phi-3.5-Mini-Instruct with LoRA-based FedPEFT. Our system consists of 15 clients in total, three of which are malicious clients attempting jailbreak attacks. We consider two data distribution scenarios: 1) IID Setting: 12 benign clients are assigned with single-domain data, either all MedQA or all MetaMathQA. 2) Non-IID Setting: 12 benign clients are evenly split between two domains (6 MedQA + 6 MetaMathQA).

Table~\ref{tb_agr} presents the utility (MedQA/MetaMathQA accuracy) and safety (AdvBench/JailbreakBench ASR) metrics across different RASs. Regarding utility, we observe that all RASs maintain or improve performance compared to the base model (initial accuracies are 56.9\% and 80.0\% for MedQA and MetaMathQA respectively) in their respective IID settings, even in the presence of malicious clients. Notably, when trained on MedQA-only data, the model significantly improves MedQA accuracy (up to 64.7\% accuracy) but declines in MetaMathQA performance (down to 71.1\%). Similarly, MetaMathQA-only training improves MetaMathQA accuracy (up to 83.6\%) with significant performance degradations in MedQA. More importantly, the results illustrate the vulnerability of RASs: Traditional RASs (Median and GeoMed) show consistently poor defense performance ($ASR > 30.7\%$) regardless of data distribution settings. While DnC and ClippedClustering effectively defend against jailbreak attacks under the MedQA setting ($ASR \leq 2\%$), they fail to maintain model safety when data is heterogeneously distributed across clients ($ASR > 80\%$).

\label{sec_agg}

\subsection{Defense with PPSA}

\label{sec_exp_ppsa}
\begin{figure}[t]
    \centering
    \includegraphics[width=\linewidth]{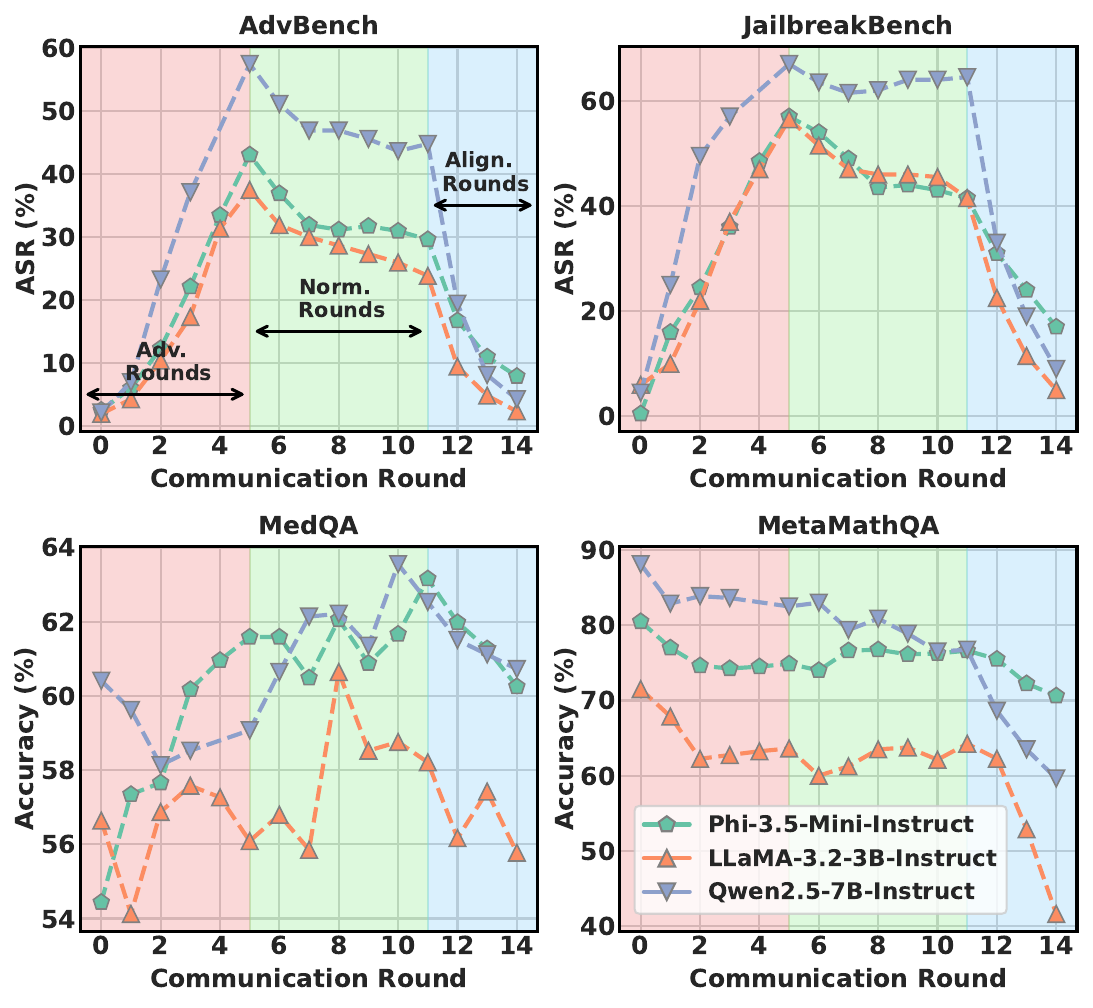}
    \caption{Evaluation of jailbreak attacks, normal fine-tuning, and alignment processes in FedPEFT across communication rounds. This study demonstrates that both jailbreak attacks and alignment can be achieved within very few communications (less than 5). However, alignment improvements may come at the expense of reduced task performance. The experiments involved fine-tuning the three PLMs on MedQA over 14 communication rounds using LoRA. During Rounds 0–5, three malicious clients were activated, nine fine-tuning clients operated between Rounds 0–11, and three alignment clients were introduced in Rounds 11–14.
    }
	\label{fig_ppsa}
\end{figure}

We further investigate the effectiveness of PPSA with LoRA fine-tuning on MedQA over just 14 communication rounds, underscoring the immediate impact of both PaaA and PPSA. In this simulation, we employ nine fine-tuning clients participating in training during rounds 0 to 10, three malicious clients active during rounds 0 to 5, and three alignment clients activated only in the final four rounds (11 to 14). As illustrated in Figure~\ref{fig_ppsa} (upper panels), in the presence of malicious clients, ASRs rapidly increase on both the AdvBench and JailbreakBench benchmarks within the first five rounds, reaching up to 57\% and 65\%, respectively. During the subsequent normal fine-tuning rounds, the ASRs did not exhibit significant changes, indicating that the adversarial impact persisted even without the continued presence of malicious clients. Eventually, in the final rounds (11 to 14), during which only alignment clients participated, the ASRs rapidly decreased to below 10\%, approximately returning to pre-attack levels.

However, this reduction comes at a cost, as alignment-focused fine-tuning often incurs an ``alignment tax''~\cite{lin-etal-2024-mitigating}, evident in the reduced utility of downstream task shown in lower panels of Figure~\ref{fig_ppsa}. For instance, in the case of LLaMA-3.2-3B-Instruct, the accuracy drops by up to 3\% on MedQA, as the model prioritizes safety over specificity. The observed accuracy drops on MetaMathQA are anticipated, given that the models are fine-tuned on MedQA. However, it is evident that during the alignment phase, this decline becomes particularly pronounced; for example, LLaMA-3.2-3B-Instruct experienced a 20\% reduction in accuracy. Thus, while PPSA effectively overcomes PaaA, it comes at the expense of performance losses in downstream tasks. This trade-off suggests that it may not be the ideal defense when preserving downstream utility is critical.

%% file: tables/setting_table.tex
\begin{table}[!t]
\caption{\small Overview of the selected PLMs and PEFT methods in our experiments. We also report trainable model parameters~(\#~TPs), total parameters~(\#~APs), and the ratio of trainable parameters.}
	\vspace*{-.4cm}
	\begin{center}
	\setlength\extrarowheight{2pt}
	\begin{adjustbox}{width=.98\linewidth,center}
        \begin{tabular}{c|c|ccc}

			  \textbf{Model}  & \textbf{Method} & \textbf{\# TPs} & \textbf{\# APs}    & \textbf{\% Params} \\
   \hline
   \hline
\multirow{3}{*}{ \textbf{LLaMA-2-7B-Chat}~\cite{touvron2023llama2}}  & LoRA & 40.0M & 6.8B  & 0.59\% \\
  & $(\text{IA})^3$ & 0.6M &  6.7B  & 0.009\% \\
    & $\text{LayerNorm}$ & 0.3M &  6.7B  & 0.004\% \\
   \hline
   \multirow{3}{*}{ \textbf{Phi-3.5-Mini-Instruct}~\cite{abdin2024phi3technicalreporthighly}} & LoRA & 29.9M & 3.9B  & 0.78\%\\
 & $(\text{IA})^3$ & 0.3M  & 3.8B  & 0.007\% \\
 & $\text{LayerNorm}$ & 0.1M &  3.8B  & 0.003\% \\
     \hline
   \multirow{3}{*}{ \textbf{LLaMA-3.2-3B-Instruct}~\cite{llama3.2}} & LoRA & 40.4M & 7.6B & 0.53\%\\
    & $(\text{IA})^3$ & 0.7M & 8.0B  & 0.009\% \\
     & $\text{LayerNorm}$ & 0.1M &  7.6B  & 0.001\% \\
  \hline
   \multirow{3}{*}{ \textbf{Qwen2.5-7B-Instruct}~\cite{qwen2.5}} & LoRA & 40.4M & 7.6B & 0.53\%\\
    & $(\text{IA})^3$ & 0.7M & 8.0B  & 0.009\% \\
     & $\text{LayerNorm}$ & 0.1M &  7.6B  & 0.001\% \\
  \hline
		\end{tabular}
	\end{adjustbox}
	\end{center}
	\label{peft_overview}
\end{table}

%% file: tables/agr.tex
\newcommand \rot{90}
\newcommand{\rotbox}[1]{\rotatebox[origin=c]{90}{#1}}

\begin{table}[t]
\caption{Breakdown of RASs against PaaA in FedPEFT: We fine-tune {Phi-3.5-Mini-Instruct} for 20 communication rounds using LoRA under IID (12 Clients with Single-Domain Data: MedQA-only or MetaMathQA-only) and Non-IID Settings (12 Clients with different domain data: 6 MedQA + 6 MetaMathQA), with 3 malicious clients. While ClippedClustering \colorbox{goodgreen}{successfully defends} against the attack in the single-domain setting with MedQA, all RASs \colorbox{badred}{fail} under the mixed-domain setting.}
    \setlength\extrarowheight{2pt}
    \begin{adjustbox}{width=\linewidth,center}
        \addtolength{\tabcolsep}{-0.5em}
        \begin{tabular}{c|l||cc|cc}
              \hline
              \multirow{2}{*}{\makecell{\textbf{FT} \\ {\scriptsize \textbf{Data}}}} & \multirow{2}{*}{\textbf{RAS}}   & \multirow{2}{*}{\makecell{\textbf{MedQA} \\ {\scriptsize \textbf{Accuracy}~(\%)}}}  & {\scriptsize \multirow{2}{*}{\makecell{\textbf{MetaMathQA} \\  \textbf{Accuracy}~(\%)}}} & {\scriptsize \multirow{2}{*}{\makecell{\textbf{AdvBench} \\ \textbf{ASR}~(\%)}}} & {\scriptsize \multirow{2}{*}{\makecell{\textbf{JailbreakBench} \\ \textbf{ASR}~(\%)}}} \\
  & &  &  &  & \\     
   \hline 
   \hline
  \rowcolor{lightgray}  \multicolumn{2}{c||}{\textbf{Base Model}}& 56.9 & 80.0 & 1.3 & 1.3 \\
\hline
 \multirow{5}{*}{\rotbox{\makecell{\textbf{MedQA} \\ \textbf{Only}}}} & \textbf{Mean}  &  64.5~\std{+7.6} &  71.1~\std{-8.9} & \cellcolor{badred} 83.5~\std{+82.2} & \cellcolor{badred} 86.0~\std{+84.7} \\
  & \textbf{Median} &  63.5~\std{+6.6} &  73.6~\std{-6.4} & \cellcolor{badred} 84.5~\std{+83.2} & \cellcolor{badred} 82.7~\std{+81.4} \\
  & \textbf{GeoMed} &  64.7~\std{+7.8} &  75.0~\std{-5.0} & \cellcolor{badred} 83.5~\std{+82.2} & \cellcolor{badred} 77.4~\std{+76.1} \\
  & \textbf{DnC} &  63.6~\std{+6.7} &  79.4~\std{-0.6} & \cellcolor{goodgreen} 1.0~\std{-0.3} & \cellcolor{goodgreen} 1.2~\std{-0.1} \\
  & \textbf{Clippedclustering} &  63.7~\std{+6.8} &  79.5~\std{-0.5} & \cellcolor{goodgreen} 0.0~\std{-1.3} & \cellcolor{goodgreen} 1.9~\std{+0.6} \\
\hline
 \multirow{5}{*}{\rotbox{\makecell{\textbf{ MetaMathQA } \\ \textbf{Only}}}}   &  \textbf{Mean} &  26.2~\std{-30.7} &  83.5~\std{+3.5} & \cellcolor{badred} 49.5~\std{+48.2} & \cellcolor{badred} 36.1~\std{+34.8} \\
  & \textbf{Median} &  27.4~\std{-29.5} &  81.5~\std{+1.5} & \cellcolor{badred} 40.5~\std{+39.2} & \cellcolor{badred} 30.7~\std{+29.4} \\
  & \textbf{GeoMed} &  31.6~\std{-25.3} &  83.6~\std{+3.6} & \cellcolor{badred} 46.5~\std{+45.2} & \cellcolor{badred} 31.7~\std{+30.4} \\
  & \textbf{DnC} &  42.4~\std{-14.5} &  82.8~\std{+2.7} & \cellcolor{badred} 85.5~\std{+84.2} & \cellcolor{badred} 85.4~\std{+84.1} \\
  & \textbf{Clippedclustering} &  53.0~\std{-3.9} &  81.0~\std{+1.0} & \cellcolor{goodgreen} 3.0~\std{+1.7} & \cellcolor{goodgreen} 3.5~\std{+2.2} \\
\hline
 \multirow{5}{*}{\rotbox{\makecell{\textbf{MedQA \&} \\ \textbf{MetaMathQA}}}}  & \textbf{Mean} &  59.5~\std{+5.0} &  82.8~\std{+2.8} & \cellcolor{badred} 77.5~\std{+76.2} & \cellcolor{badred} 77.4~\std{+76.1} \\
  & \textbf{Median} &  58.2~\std{+3.7} &  81.6~\std{+1.6} & \cellcolor{badred} 80.0~\std{+78.7} & \cellcolor{badred} 81.8~\std{+80.5} \\
  & \textbf{GeoMed} &  61.8~\std{+7.3} &  81.5~\std{+1.5} & \cellcolor{badred} 76.0~\std{+74.7} & \cellcolor{badred} 72.4~\std{+71.1} \\
  & \textbf{DnC} &  52.0~\std{-2.5} &  83.2~\std{+3.2} & \cellcolor{badred} 82.5~\std{+81.2} & \cellcolor{badred} 87.7~\std{+86.4} \\
  & \textbf{Clippedclustering} &  61.1~\std{+6.6} &  81.3~\std{+1.3} & \cellcolor{badred} 81.0~\std{+79.7} & \cellcolor{badred} 88.3~\std{+87.0} \\
\hline
        \end{tabular}
    \end{adjustbox}
    \label{tb_agr}
\end{table}

%% file: related_work.tex
\section{Related Work}

In this section, we review related work under the context of FedPEFT in Section~\ref{sec_fedpeft}, followed by fine-tuning-based jailbreak attacks and defenses during the adaptation of PLMs in Section~\ref{subsec_jb}.

\subsection{Federated Parameter-Efficient Fine-Tuning}

\label{sec_fedpeft}
Originating from the fine-tuning practices of transformer models~\cite{lester-etal-2021-power,hu2022lora}, FedPEFT is a training paradigm aiming to adapt pre-trained models based on PEFT and FL techniques, while preserving data privacy and minimizing communication costs
and computational overhead~\cite{pmlr-v232-malaviya23a}. %
Some studies have demonstrated their efficiency in training and evaluating the comparative advantages and disadvantages of different PEFT methods in terms of performance~\cite{che2023federated,zhang2024federated}, resource efficiency~\cite{sun2023fedbpt}, %
and model personalization~\cite{anonymous-acl24-arr}. %
Some other studies focus on exploring the potential of FedPEFT techniques within resource heterogeneity FL environments, such as using heterogeneous LoRA ranks and sparsity to facilitate limited and heterogeneous system capabilities of clients~\cite{su2023fedra,bai2024federated}.

Furthermore, recent work has also focused on the security aspect, which involves identifying security threats to this emerging paradigm~\cite{li2023backdoor} and developing defense mechanisms to safeguard the fine-tuning process~\cite{bi2024securingfederatedlearningnovel,ye2024emergingsafetyattackdefense}, with the objective of ensuring security and maintaining performance advantage simultaneously.

\subsection{Fine-Tuning-based Jailbreak Attacks and Defenses} 

\label{subsec_jb}
In the past few years, a branch of research has concentrated on the susceptibility of PLMs to fine-tuning-based jailbreak attacks~\cite{xu-etal-2024-comprehensive}, where malicious users exploit fine-tuning techniques to circumvent the established safety guardrails of PLMs, hereby inducing harmful behaviors. Specifically, several pioneering works~\cite{qi2024finetuning,yang2024shadow,zhan-etal-2024-removing} showed that PLMs aligned by RLHF or SFT can be jailbroken by fine-tuning with toxic data. Moreover, a line of work~\cite{leong2024devilsalikeunveilingdistinct,weiassessing,peng2024navigatingsafetylandscapemeasuring,jain2024makesbreakssafetyfinetuning} focused on accessing and analyzing how jailbreak attacks disrupt the safety guardrails and the associated implications. 

To mitigate these vulnerabilities, researchers have been actively exploring defense mechanisms across different model training and adaptation stages. 1) Alignment stage defense aims to enhance the model’s immunization ability to fine-tune by modifying the training procedure in the alignment stage~\cite{huang2024vaccine}. For instance, Vaccine~\cite{huang2024vaccine} vaccinates the model by adding embedding perturbation in the alignment stage, and RepNoise improves the robustness by enforcing the representation of the harmful data to be a random Gaussian noise. 
2) Fine-tuning stage defense focuses on maintaining the alignment knowledge while training on the downstream dataset~\cite{mukhoti2024finetuningcripplefoundationmodel,zong2024safety,huang2024lisalazysafetyalignment}. For example, LDIFS~\cite{mukhoti2024finetuningcripplefoundationmodel} introduces a regularizer to enforce the embedding of the iterate to be close to that of the aligned model. VLGuard~\cite{zong2024safety} mixes the alignment data with the user data to train in the user fine-tuning stage. Lisa~\cite{huang2024lisalazysafetyalignment} alternatively optimizes the alignment data and the user fine-tuning data and uses a proximal regularizer to enforce proximity between iterates.
3) Post-fine-tuning stage defense adopts safety alignment techniques to restore safety guardrails after malicious fine-tuning~\cite{ye2024emergingsafetyattackdefense,huang2024antidotepostfinetuningsafetyalignment,yi2024safetyrealignmentframeworksubspaceoriented}.

To our knowledge, only one preliminary study~\cite{ye2024emergingsafetyattackdefense} has investigated jailbreak attacks in federated fine-tuning settings, in which the authors proposed a PPSA based on a pipeline that automatically generates defense data and performs safety alignment on the server side. However,~\cite{ye2024emergingsafetyattackdefense} is mainly limited in the single focus on LoRA fine-tuning with a LLaMA-2 model. In contrast, our work extensively investigates several PEFT methods, PLMs, and defenses under different FL settings. Through the broader scope, we draw more generalizable conclusions about the attack's effectiveness across various scenarios, demonstrating the limitations of current defenses and emphasizing the need for advanced approaches.